\documentstyle[prl,aps,epsf,twocolumn]{revtex}

\begin{document}

\twocolumn[\hsize\textwidth\columnwidth\hsize\csname 
@twocolumnfalse\endcsname

\author{Barbara M. Terhal$^{1}$, Isaac L. Chuang$^{2}$, David P. DiVincenzo$^{3}$, Markus
Grassl$^{4}$ and John A. Smolin$^{3}$}

\title{Simulating quantum operations with mixed environments} 

 \address{\vspace*{1.2ex}
            \hspace*{0.5ex}{$^1$ Instituut voor 
Theoretische Fysica, Universiteit van Amsterdam, Valckenierstraat 65, 
1018 XE Amsterdam, and  CWI, Kruislaan 413, 1098 SJ Amsterdam, The Netherlands,
$^2$ IBM Almaden Research Center, 650 Harry Road, San Jos\'e, CA 95120,  
$^3$ IBM T.J. Watson Research Center, Yorktown Heights, NY 10598, 
$^4$ Institut f{\"u}r Algorithmen und Kognitive Systeme, Universit{\"a}t 
Karlsruhe, 76128 Karlsruhe, Germany}}

\date{\today}
\pagestyle{plain}
\pagenumbering{arabic}
\maketitle


\begin{abstract}
We study the physical resources required to implement general quantum
operations, and provide new bounds on the minimum possible size which
an environment must be in order to perform certain quantum operations.
We prove that contrary to a previous conjecture, not all quantum operations on a single-qubit can
be implemented with a single-qubit environment, even if that environment is
initially prepared in a mixed state. We show that a mixed 
single-qutrit environment is sufficient to implement of a 
special class of operations, the generalized depolarizing channels.
\end{abstract}

]



\def\be{\begin{equation}}
\def\ee{\end{equation}}
\def\bea{\begin{eqnarray}}
\def\eea{\end{eqnarray}}
\def\Adag{A^\dagger}
\def\g{\gamma}
\def\eps{\epsilon}

\newtheorem{theo}{Theorem}
\newtheorem{defi}{Definition}
\newtheorem{lem}{Lemma}
\newtheorem{exam}{Example}
\newtheorem{prop}{Property}
\newcommand{\dubbelR}{{\sf I}\kern-.12em{\sf R}}

\newcommand{\vectwoc}[2]{\left(
        \begin{array}{c}{#1}\\{#2}\end{array}\right)}
\newcommand{\mattwoc}[4]{\left[
        \begin{array}{cc}{#1}&{#2}\\{#3}&{#4}\end{array}\right]}
\newcommand{\ket}[1]{\mbox{$|#1\rangle$}}
\newcommand{\bra}[1]{\mbox{$\langle #1|$}}
\newcommand{\ba}{\begin{array}}
\newcommand{\ea}{\end{array}}
\def\>{\rangle}
\def\<{\langle}

\def\lbL{\lb\rule{0pt}{2.4ex}}
\def\lpL{\left(\rule{0pt}{2.4ex}}
\def\lb{\left[}
\def\lp{\left(}
\def\rb{\right]}
\def\rp{\right)}

\def\wa{\omega_a}
\def\wb{\omega_b}
\def\wab{\omega_{ab}}
\def\ra{\rightarrow}

\def\Ybar{\bar Y}
\def\Xbar{\bar X}
\def\Zbar{\bar Z}
\def\ep{\epsilon}

\newcommand{\mypsfig}[2]{\psfig{file=#1,#2}}

\newcommand{\myfig}[4]{
        \begin{figure}[hbtp]
        \begin{center}
        \mbox{{\mypsfig{#1}{#2}}}
        \end{center}
        \caption{#3}
        \label{fig:#4}
        \end{figure}
}


\vspace*{-4ex}

Future quantum computers may be useful in studying the behavior of
open quantum systems and the nature of decoherence \cite{misc,lloyd}.
Instead of performing real experiments on quantum systems, a single
quantum computer can be used as an efficient, multiple-purpose
simulator for a wide variety of physical systems.  In general, an
important goal of such investigations will be to understand the
effects arising from interactions between the system of interest $S$
and another quantum system $E$.  For example, a quantum computer can
be used to simulate quantum systems $S$ in thermal
equilibrium\cite{bmt_dpdv}, but such a simulation requires an
additional quantum system $E$, coupled in a particular way to $S$, to
mimic the thermal bath of the system.  In other applications involving
the simulation of nonequilibrium quantum properties, $S$ could, for a
molecules whose isomerization dynamics we wish to study, represent the
relevant conformational states, which couple to other molecules $E$
through long-range electronic dipolar interactions.  In all these
applications, we wish to to implement $E$ with the smallest quantum
resources possible, and this Letter investigates the most efficient
implementation of such quantum environments.

Suppose $S$ exists in a Hilbert space ${\cal H}_n$ of dimension $n$, and 
$E$ is in ${\cal H}_r$ of dimension $r$. It is well known that any {\em quantum
operation}\cite{schum} on ${\cal H}_n$, resulting from some interaction with 
$E$ in ${\cal H}_r$ with arbitrary $r$, can be performed by appending a state
in ${\cal H}_{n^2}$, evolving unitarily, and then tracing over ${\cal
H}_{n^2}$.  The difference between $r$ and $n^2$ can represent a significant
reduction, since $E$ can be a large bath (for
example, of harmonic oscillators), and $r$ can be infinite.

Can a general quantum operation be implemented with an environment even
smaller than $n^2$ dimensions?  Lloyd conjectured \cite{lloyd} that it is
possible to implement a general quantum operation on $k$ quantum bits (qubits)
with a $k$-qubit environment -- if one prepares the environment not in a pure
state, as is usually the case, but rather in an arbitrary {\em mixed} state.

Here we provide a specific counterexample to this conjecture for $k=1$,
although we find that at least for some operations, fewer resources are
required than was previously known.  Our counterexample is part of a class
known as the generalized depolarizing channels, for which we show that a
three-dimensional environment is sufficient for simulation.  The proof of the
counterexample is established by the technique of computing Gr\"obner bases.


Our results also address the following question: suppose a physical
system is given as a black box -- we can prepare $S$ in an arbitrary
initial state, and then measure the final state of $S$ after a fixed
evolution period.  What is the largest environment $E$ with which $S$
might have interacted in this system?  A method to completely
determine the quantum operation $\chi$ performed by this system is
known\cite{tomo}.  This work goes one step further, by showing a way
to turn knowledge about $\chi$ into bounds on the nature of $E$.

\begin{figure}
\epsfxsize=7cm
\epsfbox{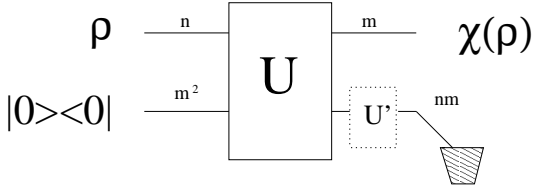}
\vspace*{3ex}
\caption{Implementation of the map $\chi$ using a pure state
environment.}
\label{fig1}
\end{figure}

We begin by summarizing the mathematical formalism of quantum operations.  The
most general transformation allowed by quantum mechanics for an initially
isolated quantum system is a linear, trace-preserving, completely positive
map.  Such a map $\chi\;\colon {\cal A}_n \rightarrow {\cal A}_m$, where
${\cal A}_n$ is the set of operators on a Hilbert space ${\cal H}_n$, 
 can be decomposed into a set of at most $nm$ $m \times n $ matrices $A_i$ \cite{choi}
(which we shall refer to as ``operation elements'') as 
\be
        \chi(\rho)=\sum_{i=1}^{nm} A_i \rho A_i^{\dagger}.
\label{first}
\ee
$n$ and $m$ are the dimensions of the input and output Hilbert spaces,
respectively.  The trace-preserving property implies that the $A_i$ obey the
constraint
\be
\sum_{i=1}^{nm} A_i^{\dagger}A_i={\bf 1}_n.
\label{uniconst}
\ee
with ${\bf 1}_n$ the identity matrix on ${\cal H}_n$. Following Choi \cite{choi}, the set of all
such maps $\chi\;\colon {\cal A}_n \rightarrow {\cal A}_m$ we call
${\bf TCP}[n,m]$. A physical implementation of these maps is
represented in Fig.~\ref{fig1}: A unitary operation on the state $\rho
\otimes \ket{0} \bra{0}$ (where $\ket{0}$ represents some pure state in
an $m^2$-dimensional environment) is performed and then $nm$ ``degrees
of freedom'' are traced out:
\be
        \chi(\rho)= \sum_{k=1}^{nm}  
                \<e_k|\, U \lbL{ \rho \otimes |0\>\<0| }\rb U^\dagger \,|e_k\>.
\label{upure}
\ee
Here $\{\ket{e_k}\}_{k=1}^{nm}$ is a set of basis vectors for ${\cal H}_{nm}$.
As there are at most $nm$ operation elements, it follows that one can
implement any map in ${\bf TCP}[n,m]$ with an environment of dimension
$m^2$. To determine the dimension of the parameter space of ${\bf TCP}[n,m]$
we note that the map $\chi$ does not uniquely determine the set
$\{A_i\}_{i=1}^{nm}$. Any set of $m \times n$ matrices $\{B_i\}_{i=1}^{nm}$
and $\{A_j\}_{j=1}^{nm}$ that are related by a unitary transformation
\be
        B_i=\sum_j^{nm} U'_{ij} A_j 
\label{unifree}
\ee
implement the same map $\chi$. This freedom corresponds to a unitary 
rotation $U'$ (see Fig.~\ref{fig1}) of the environment qubits after the completion of the interaction 
$U$. It is shown in \cite{schum} that this unitary equivalence is the 
only freedom in the choice for the set of operators $\{A_i\}_{i=1}^{nm}$.
 
The dimension of the parameter space of all maps in ${\bf TCP}[n,m]$ that can be implemented with a $d$-dimensional environment will therefore be 
\be
D_{\mbox{\scriptsize pure},d}^{n \rightarrow m} 
        =\!\!\overbrace{2 n^2 d}^{\mbox{\scriptsize 
        parameters in }\{A_i\}}- 
\overbrace{(nd/m)^2}^{\mbox{\scriptsize unitary freedom}}-\overbrace{n^2}^{\mbox{
\scriptsize constraint }(\ref{uniconst})} 
\label{par_pure}
\ee
where $d$ is such that $m$ divides $nd$. Thus we have $D_{{\bf TCP}[n,m]}=D_{\mbox{\scriptsize pure},m^2}^{n \rightarrow m}=n^2(m^2-1)$. 

In a more general physical implementation, however, the initial state of the
environment can be an arbitrary density matrix.  Consider the set of
completely positive trace-preserving linear maps $\chi\;\colon {\cal A}_n \rightarrow
{\cal A}_m$ that are implemented by an environment that is initially in some
$d$-dimensional density matrix. We call this set $S_{\mbox{\scriptsize
mix}}[d,n,m]$. The action on the input state $\rho$ is
\be
        \chi(\rho)= \sum_{j=1}^d \lambda_j \sum_{k=1}^{dn/m}  
                \<e_k|\, U \lbL{ \rho \otimes |j\>\<j| }\rb U^\dagger\, |e_k\>
\,,
\ee
where $\{\lambda_j,\ket{j}\}_{j=1}^d$ are now the eigenvalues and eigenvectors
of the mixed environment state. We identify a set of $m \times n$ matrices $\{
A_{jk} \}_{j=1,k=1}^{d,dn/m}$ in the representation of Eq. (\ref{first}):
\be
        A_{jk}=\sqrt{\lambda_{j}} \bra{e_k} U \ket{j}.
\label{u_in_a}
\ee
Unitarity implies that these matrices are constrained, 
\be
\sum_k A_{ik}^{\dagger}A_{jk}=\delta_{ij}\lambda_i {\bf 1}_n.
\label{uninewcon}
\ee
There is a residual unitary freedom in choosing the set of matrices
$\{A_{jk}\}_{j=1,k=1}^{d,nd/m}$. The set $\{B_{jm}\}_{j=1,m=1}^{d,nd/m}$ with
$B_{jm}= \sum_k U'_{mk} A_{jk}$, where the $d\, n/{m}$-dimensional unitary matrix $U'$ does not depend on
the label $j$, implements the same quantum operation and also obeys constraint
(\ref{uninewcon}). As before, this freedom corresponds to a unitary
transformation on the environment after the completion of the operation. The
dimension of the parameter space of $S_{\mbox{\scriptsize mix}}[d,n,m]$ can be
bounded:
\be
\ba{c}
D_{\mbox{\scriptsize pure},d}^{n \rightarrow m} \leq D_{\mbox{\scriptsize mix},d}^{n \rightarrow m} \leq
D_{\mbox{\scriptsize pure},d^2}^{n \rightarrow m}. 
\ea
\label{mix}
\ee
The upper bound is given by the fact that one can always simulate a 
$d$-dimensional mixed environment with a $d^2$-dimensional pure environment.

From Eq.(\ref{par_pure}) and Eq.(\ref{mix}) it follows that an environment of dimension $d < m$ cannot be used to implement ${\it
all}$ maps in ${\bf TCP}[n,m]$. In fact a large set of maps, the extremal maps
in ${\bf TCP}[n,m]$, cannot be simulated with $d < m$.  A map $\chi$ that is
decomposable in $m$ or fewer linearly-independent operation elements is
extremal \cite{choi} in ${\bf TCP}[n,m]$.  These maps can be implemented with
a pure-state environment of dimension $m$; moreover, we prove that
there does not exist a more efficient implementation of these maps using a
mixed-state environment:

Extremality implies that the map $\chi$ cannot be written as a convex
combination of linearly independent maps $\chi^i$ that each have
operation elements $\{A_j^i\}$ for which $\sum_j {A_j^i}^{\dagger}A_j^i={\bf
1}_n$ for each $i$. This ensures that only one of the eigenvalues in
constraint (\ref{uninewcon}) is non-zero, but this in fact corresponds to a
pure-state environment of dimension $m$. An example of such an extremal map is
a von Neumann measurement on a $n$-dimensional system. The set of projection
operators $\{P_i\}_{i=1}^n$ can be implemented minimally by using an 
$n$-dimensional pure state.

We now turn to the question of whether all maps in ${\bf TCP}[n,m]$ can be
implemented with $d=m$. Note that our parameter count does not exclude this. In the following, we restrict ourselves to the case $n=m=2$. We
study which maps can be implemented using a single-qubit environment and
provide a proof that a particular qubit channel, the two-Pauli channel, cannot be implemented
in this way.

We consider a special set of maps, the generalized depolarizing channels \cite{2pauli},
which are described by the set $\{(\eps_i, A_i)\}_{i=1}^4$ where
\be
\chi (\rho)=\sum_i \eps_i A_i \rho A_i^{\dagger}.
\label{gendep}
\ee
such that $\epsilon_1+\epsilon_2+\epsilon_3+\epsilon_4=1$ and the 
operators $A_i$ are given by
$A_1={\bf 1}_2$, $A_2=\sigma_x$, $A_3=\sigma_y$, $A_4=\sigma_z$. One can represent this family of maps geometrically as a tetrahedron, which is
embedded in a cube with vertices at $(1,-1,-1)$, $(-1,1,-1)$, $(1,1,1)$ and
$(-1,-1,1)$. The transformation that relates the parameters
$\epsilon_1,\epsilon_2,\epsilon_3,\epsilon_4$ to the $(x,y,z)$ coordinates
is given by $x=\epsilon_1+\epsilon_2-\epsilon_3-\epsilon_4$,
$y=\epsilon_1-\epsilon_2+\epsilon_3-\epsilon_4$, and
$z=\epsilon_1-\epsilon_2-\epsilon_3+\epsilon_4$. The vertices of the tetrahedron correspond to a single-operator map. Its edges
are two-operator maps, the four faces represent all three-operator maps, and
the points in the interior of the tetrahedron are all the four-operator maps
of Eq. (\ref{gendep}).

A computer search suggests that only a subset of these maps can
be simulated by using a qubit environment. For this subset we are able
to construct an explicit qubit solution. At web address \cite{web} one
can find pictures of the three-dimensional volume that is described by
the solution set and a picture of the solution set as generated by the
computer search.  The computer work also suggests that the dimension 
of $S_{\mbox{\scriptsize mix}}[2,2,2]$ is equal 
the upper bound of Eq.(\ref{mix}), namely 
$D_{\mbox{\scriptsize pure},4}^{2 \rightarrow 2}=12$. We find this by randomly sampling in the space of all superoperators, that is, we choose random orthonormal vectors that make up the 
columns of the unitary matrix $U$ of Eq. (\ref{upure}); a finite 
percentage could be implemented with a qubit environment. Thus there is 
enough ``room'' for a solution, but it is not in the right place, as we 
will see.

This solution is constructed in the following way.  We start with the center
of mass of the tetrahedron, the point
$(\eps_1,\eps_2,\eps_3,\eps_4)=(1/4,1/4,1/4,1/4)$. This channel has the
property that it maps every input state $\rho$ onto $\frac{1}{2}{\bf 1}_2$. It
can thus be easily implemented by performing a SWAP gate on a environment
qubit that is initially in the $\frac{1}{2}{\bf 1}_2$ state and the input
qubit. The SWAP gate on two registers $\ket{a}\ket{b}$ gives
$\ket{b}\ket{a}$. Then one considers the line that departs from a vertex, say
the point $(\eps_1,\eps_2, \eps_3,\eps_4)=(1,0,0,0)$, and goes through the
center of mass. This one-dimensional set of channels is characterized by
$\eps_2=\eps_3=\eps_4$ and represents the regular depolarizing channel
\cite{2pauli}. Performing a $\sqrt[m]{\mbox{SWAP}}$ on a $\frac{1}{2}{\bf
1}_2$ environment and the input qubit implements these channels, up to
$\eps_1=1/4$. The integer $m$ is related to the $\epsilon$ parameters
by $\epsilon_2=\epsilon_3=\epsilon_4=\sin^2(\frac{\pi}{2m})/4$. One extra step of generalization gives us an even larger set of
channels. The unitary matrix is a somewhat generalized form of
$\sqrt[m]{\mbox{SWAP}}$,
\be
U=\left(
\ba{lccr}
e^{i \theta }\cos \phi_1 & 0 & 0 & i e^{i \theta }  \sin \phi_1 \\
0 & \cos\phi_2 & i\sin\phi_2 & 0 \\
0 & i\sin\phi_2 & \cos \phi_2 & 0 \\
 i e^{i \theta } \sin \phi_1 & 0 & 0 & e^{i \theta }\cos \phi_1
\ea
\right),
\label{uniswap_gen}
\ee
and the environment is again prepared in state $\frac{1}{2}{\bf 1}_2$. We can
determine the operation elements and express these as linear (unitary)
combinations of the Pauli matrices. This leads to an expression of the
parameters $\epsilon_i$ in terms of $(\theta,\phi_1,\phi_2)\in
[0,2\pi]\times[0,2\pi]\times[0,2\pi]$:
\be
\ba{l}
\epsilon_1= \frac{1}{4}(\cos^2 \phi_1+ \cos^2 \phi_2+
2 \cos \phi_1 \cos \phi_2 \cos \theta), \\
\epsilon_2=\frac{1}{4}(\sin^2 \phi_1+ \sin^2 \phi_2+
2 \sin \phi_1 \sin \phi_2 \cos \theta), \\
\epsilon_3=\frac{1}{4}(\sin^2 \phi_1+ \sin^2 \phi_2-
2 \sin \phi_1 \sin \phi_2 \cos \theta),\\
\epsilon_4=\frac{1}{4}(\cos^2 \phi_1+ \cos^2 \phi_2-
2 \cos \phi_1 \cos \phi_2 \cos \theta). 
\ea
\label{para}
\ee



We now turn to another set of maps, the two-Pauli channel, which is given by
the 
three operators
\be
A_1={\bf 1}_2\sqrt{x}, A_2=\sigma_x\sqrt{(1-x)/2},
        A_3=i\sigma_y\sqrt{(1-x)/2}
\,.
\ee
We will prove that for $0< x < 1$, there is no qubit
environment which simulates this channel. For $x=0$ or $x=1$ there is a
two-dimensional environment that can simulate the channel as the channel has
two operation elements when $x=0$ and only one operator when $x=1$.

Any unitary linear combination of the $A_1,A_2$ and $A_3$ may be
written as
\be
B_k = \left(\ba{cc} b_k\sqrt{x} & (c_k-a_k)\sqrt{\frac{1}{2}(1-x)} \\
(c_k+a_k)\sqrt{\frac{1}{2}(1-x)} & b_k\sqrt{x}\ea\right)
\label{transformAB}
\,,
\ee
with appropriate constraints resulting from unitarity on the coefficients
$a_k,b_k,c_k$. This new set of operators $\{B_k\}_{k=1}^4$ will implement the
same channel due to Eq.(\ref{unifree}). Furthermore, these operators $B_k$ are
constrained through Eq.~(\ref{uninewcon}).  For notational convenience, we
define
\be
\ba{cc}
|u_0\> = \frac{1}{\sqrt{2}}\vectwoc{a_0+c_0}{a_1+c_1}, & |u_1\> = \frac{1}{\sqrt{2}}\vectwoc{a_2+c_2}{a_3+c_3}, \\
|w_0\> = \frac{1}{\sqrt{2}}\vectwoc{c_0-a_0}{c_1-a_1}, & 
|w_1\> = \frac{1}{\sqrt{2}}\vectwoc{c_2-a_2}{c_3-a_3}, \\
|v_0\> = \vectwoc{b_0}{b_1}, & |v_1\> = \vectwoc{b_2}{b_3}. \\
\ea
\ee

Using the assumption $0\ne x \ne 1$ and by linearly combining all the
equations we obtain:
\bea
        \<v_0 |w_0\> + \<u_0 |v_0\> & =: g_{1}  =& 0 
\label{g1}
\\      \<v_1 |w_1\> + \<u_1 |v_1\> & =: g_{2}  =& 0 
\\      \<v_0 |w_1\> + \<u_0 |v_1\> & =: g_{3}  =& 0 
\\      \<w_0 |v_1\> + \<v_0 |u_1\> & =: g_{4}  =& 0 
\\      \<u_0 |u_0\> - \<w_0 |w_0\> & =: g_{5}  =& 0 
\\      \<u_1 |u_1\> - \<w_1 |w_1\> & =: g_{6}  =& 0 
\\      \<u_0|u_0\> + \<u_1|u_1\>-1 & =: g_{7}  =& 0 
\\      \<v_0|v_0\> + \<v_1|v_1\>-1 & =: g_{8}  =& 0 
\\      \<u_0 |v_0\> + \<u_1 |v_1\> & =: g_{9}  =& 0 
\\      \<u_0 |w_0\> + \<u_1 |w_1\> & =: g_{10} =& 0 
\\      \<u_0 |u_1\> - \<w_0 |w_1\> & =: g_{11} =& 0 
\label{g11}
\eea

Writing each of the coefficients $a_k$, $b_k$, and $c_k$ in the form
$x_{j}+i x_{j+1}$ (where $i^2=-1$), we get a system of polynomial
equations $\mbox{Re}(g_{1})=\mbox{Im}(g_1)=\ldots=\mbox{Im}(g_{11})=0$, where $\mbox{Re}(g_k)$ and $\mbox{Im}(g_k)$ are polynomials in the variables $x_{1},\ldots,x_{24}$ with real
coefficients. To show that this system of equations has no solution we
make use of Gr{\"o}bner bases (see e.\,g.~\cite{CLOS92}). The
computation of a Gr{\"o}bner basis with Buchberger's algorithm
generalizes the Euclidean algorithm to compute the greatest common
divisor (GCD) of univariate polynomials $p_1(x)$ and $p_2(x)$. In that
case, the GCD $g(x)$ can be written as a ``linear'' combination
$g(x)=f_1(x) p_1(x) + f_2(x) p_2(x).$
The two univariate polynomials $p_1$ and $p_2$ have a common root if and only
if their GCD is non-trivial, i.\,e., $g(x)\ne 1$.

For multivariate polynomials, a common solution exists iff the Gr{\"o}bner
basis of the ideal generated by them is non-trivial, i.\,e., does not contain
a constant. In our case, using the computer algebra system {\sc Magma}
\cite{Magma} we have shown that there exist polynomials $f_1,\ldots,f_{11}$
such that
$\sum_{j=1}^{11} f_j(x_1,\ldots,x_{24}) g_j(x_1,\ldots,x_{24})=1$,
i.\,e., the Gr{\"o}bner basis contains $1$ and there is no solution of
the equations (\ref{g1})--(\ref{g11}). $\Box$


Despite the above proof, it turns out that the class of channels we
have been studying do not require a two qubit environment ($d=4$) for their
simulation; a mixed {\em qutrit} ($d=3$) suffices.  For generalized
depolarizing channels, there will be nine operators,
$\{A_{ij}\}_{i,j=1}^3$.  We set one eigenvalue $\lambda_3=0$ and thus
$A_{31}=A_{32}=A_{33}=0$.  If $\epsilon_1
\epsilon_2 \geq \epsilon_3 
\epsilon_4$ the solution is $A_{11}=0$, $A_{21}=
\sqrt{\epsilon_2-{\epsilon_3 \epsilon_4}/{\epsilon_1}} \sigma_x$,
$A_{12}=\sqrt{\epsilon_3}\sigma_z$,
$A_{22}=\sqrt{\epsilon_4}\sigma_y$, $A_{13}= \sqrt{\epsilon_1}{\bf
1}_2$, and $A_{23}=-i\sqrt{{\epsilon_3
\epsilon_4}/{\epsilon_1}}\sigma_x$.  Otherwise, we take $A_{11}=0$,
$A_{21}= \sqrt{\epsilon_4-{\epsilon_1 \epsilon_2}/{\epsilon_3}}
\sigma_y$, $A_{12}=\sqrt{\epsilon_1}{\bf 1}_2$,
$A_{22}=\sqrt{\epsilon_2}\sigma_x$,
$A_{13}=\sqrt{\epsilon_3}\sigma_z$, and
$A_{23}=i\sqrt{{\epsilon_1 \epsilon_2}/{\epsilon_3}}\sigma_y$.
One can check that this set implements any generalized depolarizing
channel and 
satisfies Eq. (\ref{uninewcon}). 


On the basis of the computer work we conjecture that {\em any} map in ${\bf
TCP}[2,2]$ can be simulated with a qutrit environment.  Also, the numerics
suggest that one can always set one eigenvalue to zero.  Furthermore, we have
some numerical evidence that channels that have three linearly independent
operation elements can never be simulated with a qubit environment.


Our results provide new bounds on the size of an environment needed to
simulate certain quantum operations on single qubits.  However, we have only
addressed simple mappings on the smallest input space.  Many questions now
arise: how do these results generalize to mappings on $n$-dimensional systems?
A relevant scenario might be $n$ uses of the generalized depolarizing channel,
where the environment can be shared between the channels.  In such a case,
might a qubit environment per channel suffice for large $n$?  A nice
extension of the generalized depolarizing channels are the channels
that are defined with the Heisenberg group elements \cite{heis}. These
channels on $n$-dimensional inputs are mixtures of a set of $n^2$
unitary matrices $U(i,j)$. However, it is not straightforward to construct solutions, as in the
qutrit case, for a general ``Heisenberg channel,'' and we have no
insight at the moment of what gain one can get by using mixed states
here.  The questions we have formulated also apply to the construction
of generalized measurements: how large an environment is needed for the minimal-size construction of arbitrary generalized measurements on an $n$-dimensional system?  We
hope our results and the questions they motivate will be useful in
future quantum computing applications, and provide fundamental
insights into the often strange properties of quantum systems.



\end{document}